\documentclass[letterpaper]{emulateapj} 
\usepackage{apjfonts}                   

\usepackage{epsfig,graphicx,latexsym,amsmath,amssymb}
\usepackage{natbib}
\usepackage{hyperref}
\usepackage{mathrsfs}
\usepackage{lastpage}

\def\msun{M_\odot}
\def\rsun{R_\odot}
\def\mbh{M_\bullet}
\def\t2b{t_{\rm 2b}}

\def\kms{{\rm km~s^{-1}}}

\bibpunct[,]{(}{)}{;}{a}{}{,}

\begin{document}

\author{
Pau Amaro-Seoane\altaffilmark{1}\thanks{e-mail: Pau.Amaro-Seoane@aei.mpg.de} \&
Xian Chen\altaffilmark{1}\thanks{e-mail: Xian.Chen@aei.mpg.de}
}

\altaffiltext{1}{Max Planck Institut f\"ur Gravitationsphysik
(Albert-Einstein-Institut), D-14476 Potsdam, Germany.}

\date{\today}

\label{firstpage}

\title{The fragmenting past of the disk at the Galactic Center :\\
The culprit for the missing red giants}

\begin{abstract}

Since 1996 we have known that the Galactic Center (GC) {displays a core-like
distribution of red giant branch (RGB) stars} starting at $\sim 10\arcsec$,
which poses a theoretical problem, because the GC should have formed a
segregated cusp of old stars.  This issue has been addressed invoking stellar
collisions, massive black hole binaries, and infalling star clusters, which can
explain it to some extent.  Another observational fact, key to the work
presented here, is the presence of a stellar disk at the GC. We postulate that
the reason for the missing stars in the RGB is closely intertwined with the
disk formation, which initially was gaseous and went through a fragmentation
phase to form the stars.  Using simple analytical estimates, we prove that
during fragmentation the disk developed regions with densities much higher than
a homogeneous gaseous disk, i.e. ``clumps'', which were optically thick, and
hence contracted slowly. Stars in the GC interacted with them and in the case
of RGB stars, the clumps were dense enough to totally remove their outer
envelopes after a relatively low number of impacts. Giant stars in the
horizontal branch (HB), however, have much denser envelopes.  Hence, the
fragmentation phase of the disk must have had a lower impact in their
distribution, because it was more difficult to remove their envelopes.  We
predict that future deeper observations of the GC should reveal less depletion
of HB stars and that the released dense cores of RGB stars will still be
populating the GC.

\end{abstract}


\maketitle

\section{Introduction}
\label{sec:GC}

The observations of the inner {0.5 pc ($12\arcsec$)} of the GC has led in
recent years to interesting and challenging discoveries that cannot be fully
addressed in the context of standard two-body relaxation theory \citep[for a
general summary about the GC, see e.g.][]{gen10}. On the one hand,
{\citet{buc09,do09}} discovered a spherical core of RGs with a flat surface
density profile.  {If} these RGs  trace an underlying old stellar population
{(of $\sim10^9$ years)}, the total mass of the old stars might be
$\sim10^5~\msun$ {\citep{mer10}}.  Moreover,
\citet{LevinBeloborodov03,tan06,pau06,lu09,bar10} unveiled the presence of a
mildly thick ($H/R\simeq0.1$, with $H$ the height and $R$ the radius) {and
young (2--7 Myr) stellar disk}, of about $100$ Wolf-Rayet (WR) and O-type stars
in near-circular orbits ($e<0.4$). The disk has a total mass of
$\sim10^{4}~\msun$ and a surface density profile of $\Sigma_d(R)\propto
R^{\,-2}$. The inner and outer edges of the disk are approximately at $R_{\rm
in}\simeq0.04$ pc and $R_{\rm out}\simeq0.5$ pc.  There is also an indication
for a second disk, with more eccentric stellar orbits ($e>0.6$) and smaller
disk mass ($<5\times10^3~\msun$), inclined by about $115^{\circ}$ relative to
the first one, and with a contrary rotation. However, the existence of this
second disk is still in debate {\citep{pau06,lu09,bar09}}.

The problem of the missing RGs has been addressed by a number of different
authors whose approaches can be divided into three general scenarios: (i) along
with the discovery of the missing stars in the RGB, \cite{Genzel1996} suggested
the interpretation that this could be attributed to stellar collisions due to
the extreme stellar densities reached in the GC. This idea has been explored
extensively in the works of \cite{dav98,ale99,bai99,dale09}, but it cannot
fully explain the observations; (ii) it has also been hypothesized that a
massive black hole binary could scour out a core in the GC via three-body
slingshots \citep{bau06,PZ2006,mat07,loc08,gua12}, but in order to reproduce a
core as large as what is observed, the mass of the secondary MBH at the GC
should be at least $\sim 10^5~\msun$. This would imply that the Milky Way
recently had a major merger, ruled out by current observations
\citep[e.g.][]{hansen03,yu03,chen13}; (iii) infalling clusters towards the GC
could also steepen the density profile outside $10\arcsec$, making the inner
$10\arcsec$ like a core \citep{kim03,ern09,ant12}, but strong mass segregation
can rebuild the cusp in the MW in about 1/4 of the relaxation time
\citep{AH2009,PAS2010,ASP2011}. Hence, this argument would require a steady
inflow of a cluster roughly every $10^7$ years to avoid cusp regrowth.

In this article we propose a simple, new scenario in which the depletion of RGs
is merely a consequence of the natural fragmentation phase that the gaseous
disk experienced. We prove that the regions of overdensity in the star-forming
disk {could have} removed the envelope of stars in the RGB after a rather
low number of crossings through the disk. The exact number depends on effects
of non-linearity that cannot be addressed in our simple analytical model.  In
section~\ref{sec.formation} we introduce the formation of overdensity regions
in the star-forming disk and the conditions for them to annul the envelope of
RGB stars. In~\ref{sec.number} we derive the mean number of crossing times that
a star will hit one of the clumps in the disk depending on its orbital
parameters and in~\ref{sec.impact} the net effect on the {clumps}.  We
summarize our findings in section~\ref{sec.results} as well as the main
implications.

\section{Formation of clumps in the gaseous disk and envelope removal criterion}
\label{sec.formation}

The {\it in situ} star formation model suggests that the disk of WR/OB giant
stars formed 2--7 Myrs ago in an accretion disk around the central MBH
\citep[][]{LevinBeloborodov03,GenzelEtAl2003}. To become self-gravitating and
trigger star formation, the disk initially should have had at least
$10^4~\msun$ of gas, and could have been as massive as $10^5\msun$
\citep{nay05}.  When a RG crosses the gaseous disk with a relative velocity
$v_*$, only that part of the envelope with a surface density lower than

\begin{equation} \label{eqn:sigstar}
\Sigma_*\simeq\frac{v_*}{\sqrt{Gm_*/r_*}}\Sigma_d
\end{equation}

\noindent
will be stripped off the RG by the disk because of the momentum imparted to
that section of the RG \citep{arm96}. In the above equation, $\Sigma_d$ denotes
the surface density of the disk where the impact happens, and $m_*$ and $r_*$
are the mass and radius of the RG, so that $\sqrt{Gm_*/r_*}$ represents the escape velocity
from the RG calculated at its surface. The reason why we use the value of the
escape velocity here and not at deeper radii in the RG is that the density of a
{\em homogeneous} disk,

\begin{equation}
\Sigma_d\sim \frac{10^4\msun}{\left(0.1~{\rm pc}\right)^2}\sim 10^6~\msun~{\rm pc^{-2}}\sim200~{\rm g~cm^{-2}}, \label{eqn:sigd}
\end{equation}

\noindent
is so low that when the RG crosses the disk, it will be barely scratched, i.e.
only material at the surface will be removed from it. For example, an impact at
a distance $0.1$ pc from the central MBH of mass $\mbh\simeq4\times10^6~\msun$
has a relative velocity of $v_*\sim400~\kms$. By comparing $\Sigma_*$ from
Equation~(\ref{eqn:sigstar}) and the RG model from \citet{arm96} for
$m_*\sim1~\msun$ and $r_*\sim100~\rsun$, less than $\sim10^{-7}~\msun$ of the
RG envelope will be lost due to the impact. Such a gaseous disk will not induce
any noticeable change in the structure of the RG.  Only {more massive
disks}, $\ga 10^5~\msun$, and long-lived in the gaseous phase, $\ga 10^7$ yrs,
can lead to a more efficient depletion of the envelope\footnote{As in the work
of \cite{davies13}, private communication}, but these numbers strongly
contradict current observations \citep{nay05,pau06}.

Because the disk itself is too tenuous to strip the entire envelope of any RG
flying through it, we postulate that the regions of overdensity that
progressively form in the disk, referred to as ``clumps'', are dense enough to
efficiently remove it completely and release the inner compact core of the RGs.
This depletion of RGs leads to their flat spatial distribution {\em and}
implicates the existence of a similar number of dense cores within the same
volume.

During fragmentation, a clump must satisfy the Jeans criterion to become
self-gravitating, that is, if its radius is $R_c$, the initial diameter must be
comparable to the Jeans length, i.e. $2R_c \sim \lambda_J\simeq
c_s/(G\rho)^{1/2}$, where $\rho$ is the local gas density, and $c_s$ the
effective sound speed.
Using $M_c\simeq\rho R_c^3$ and $c_s\simeq H\sqrt{G\mbh/R^3}$ in hydrostatic
equilibrium, we can now link the properties of the clump, its mass $M_c$ and
radius $R_c$, with the scale height $H$ of the disk and the distance $R$ to
SgrA$^*$,

\begin{equation}
\frac{R_c}{R}\simeq\frac{4M_c}{\mbh}\left(\frac{R}{H}\right)^2
\simeq10^{-2}\left(\frac{M_c}{10^2~\msun}\right)
\left(\frac{H/R}{0.1}\right)^{-2},\label{eqn:rc}
\end{equation}

\noindent
From the last equation we can derive the volume density
$\rho_c$ and surface density $\Sigma_c$ for the clumps,

\begin{align}
\rho_c&\simeq\rho\simeq\frac{M_c}{R_c^3}\simeq\frac{\mbh}{64R^3}\left(\frac{\mbh}{M_c}\right)^2\left(\frac{H}{R}\right)^{6}\nonumber \\
&\simeq10^{-11}{\rm g~cm^{-3}}\left(\frac{M_c}{10^2\msun}\right)^{-2}\left(\frac{H}{0.1R}\right)^6\left(\frac{R}{0.1{\rm pc}}\right)^{-3},\label{eqn:rhoc}\\
\Sigma_c&\simeq\rho_c R_c\simeq\frac{\mbh}{16R^2}\left(\frac{\mbh}{M_c}\right)\left(\frac{H}{R}\right)^{4} \nonumber\\
&\simeq 2\cdot\,10^{4}{\rm g~cm^{-2}}\left(\frac{M_c}{10^2\msun}\right)^{-1}\left(\frac{H}{0.1R}\right)^4\left(\frac{R}{0.1{\rm pc}}\right)^{-2}.\label{eqn:sigc}
\end{align}

\noindent
The stars in the disk are mainly O/WR, which have been observationally
constrained to have masses ranging between $64-128\,M_{\odot}$ \citep{zin07},
so in the following we adopt $M_c=10^2~\msun$ as the fiducial value.  We take
$H/R=0.1$ as the thickness in view of the current observations of the disk at
the GC.  Then from Equations (\ref{eqn:sigd}) and (\ref{eqn:sigc}), we can see
that a clump is typically $\sim 10^{2}$ more efficient in destroying RGs than
its analogue in an homogeneous gaseous disk.

We note that the argument that led to Equation (\ref{eqn:rc}) at the same time
ensures that the clumps will withstand the tidal forces arising from the MBH,
because the Roche radius,
$R(M_c/\mbh)^{1/3}\simeq(R/34)[M_c/(10^2\msun)]^{1/3}$, is about three times
larger than $R_c$ for an $100\,M_{\odot}$ clump.

When a clump collides with a RG of mass $m_*\simeq1~\msun$ and radius
$r_*\simeq150~\rsun$, at a relative velocity comparable to the orbital velocity
of the clump $v_c\simeq 400 [R/(0.1~{\rm pc})]^{-1/2}~\kms$, the amount of mass
stripped off from the star is

\begin{align}
M_{\rm loss} \sim &10^{-5}~\msun\frac{v_c}{\sqrt{Gm_*/r_*}}
          \left(\frac{\Sigma_c}{10^{4}~{\rm g~cm^{-2}}}\right) \nonumber \\
 \sim   &10^{-4.6}~\msun~\left(\frac{M_c}{10^2\msun}\right)^{-1}\left(\frac{R}{0.1~{\rm pc}}\right)^{-5/2}.
\end{align}

\noindent
The first line was derived by \cite{arm96} numerically, and in the second line
we have used $\Sigma_c$ from Equation~(\ref{eqn:sigc}) for scaling.

Successive impacts will remove {\em even more efficiently} the outer layer of the RG.
This is so, because the density gradient of the RG decreases (see equation 9 of \citealt{arm96}
or \citealt{TheBook}):
The enclosed mass is reduced, and the polytropic constant increases. The envelope
therefore expands to even larger radii (see upper panel of Figure 7 in \citealt{arm96}).
The timescale for the expansion is the convective time, much shorter than the orbital
period of the star -- The RG has achieved hydrostatic equilibrium much before the next impact.
To account for this effect, we assume that the $n$th impact strips a
mass of $f_{\rm loss}^{n-1}M_{\rm loss}$ from the RG, where $f_{\rm loss}>1$.
After $n$ impacts, the RG has lost a total mass of $M_{\rm loss} (f_{\rm
loss}^n-1)/(f_{\rm loss}-1)$.  In order to {\em totally} lose the envelope, we
have to equate

\begin{equation}
M_{\rm loss} \frac{f_{\rm loss}^n-1}{f_{\rm loss}-1}=M_{\rm env} \sim 0.5\,M_{\odot},
\label{eq.}
\end{equation}

\noindent
where $M_{\rm env}$ is the mass in the envelope, and so

\begin{align}
n_{\rm loss}\simeq&\frac{1}{\ln f_{\rm loss}} \left[10+\ln(f_{\rm loss}-1)+\ln\left(\frac{M_c}{10^2\msun}\right)
\right.\nonumber\\
&\left.+2.5\ln\left(\frac{R}{0.1~{\rm pc}}\right) \right].
\end{align}

\noindent
For a RG of size $r_*\simeq150~\rsun$, the typical value of $f_{\rm loss}$ is
$2$ \citep{arm96}. Hence, it takes about $14$ impacts with clumps of
$M_c\sim10^2~\msun$ located at $R\sim0.1$ pc to completely remove the RG
envelope.  The corresponding $n_{\rm loss}$ will increase to 80 (530) if we
assume $f_{\rm loss}=1.1$ (1.01).  We note that for smaller but more common
RGs, such as those at the base of the RG branch, $f_{\rm
loss}<2$ is more likely.

\section{Number of interactions with clumps}
\label{sec.number}

We now estimate the number of impacts that a RG experiences during successive
passages through the fragmenting accretion disk. At a given moment,
suppose the disk has a total of $N$ clumps. The eccentricities of these
clumps, as we saw in Section~\ref{sec:GC}, are not zero, but range between
$0.1-0.4$, ensuring a covering of the disk surface by a fraction of
$N(R_c/R)^2$ for an infalling RG whose {\em velocity vector} is perpendicular
to the disk plane. Such a RG with semimajor axis $a\la10\arcsec$ and period
$P(a)\simeq10^{3.2}~(a/0.1~{\rm pc})^{3/2}~{\rm yrs}$, will collide with clumps
at a rate $\Gamma\sim 2N(R_c/R)^2/P(a)$. Any RG on such an orbit will interact
with clumps for a time scale comparable with the fragmentation phase of the
disk, $t_{\rm frag}$. The exact value of this time depends strongly on the
initial conditions, but also on the cooling function and other variables
\citep{nay07,WardleYusef-Zadeh2008,Bon08,map12,As13}. Notwithstanding, we note
that our model does not rely on $t_{\rm frag}$: Whatever its value is, a {\em
total} number of at least $N_c\sim10^2$ clumps with $M_c\sim10^2~\msun$ will
have formed if we want to match the observed number of WR/O stars in the GC
stellar disk. Consequently, at any given moment, the disk will harbor $N\sim
N_c(t_c/t_{\rm frag})$ clumps, where we have introduced $t_c$, the lifetime of
a clump, whose value is derived later in this section. The total number of
perpendicular collisions during $t_{\rm frag}$, $n_\perp$, can be estimated to be

\begin{align}
n_\perp&\sim \Gamma t_{\rm frag}
\sim N_c\left[\frac{2t_c}{P(a)}\right]\left(\frac{R_c}{R}\right)^2.
\end{align}

\noindent
As mentioned before, there is no dependence on $t_{\rm frag}$ itself.
On the other hand, if the orbital plane of RG is coplanar with the
disk, the path of the RG covered inside the disk will be longer than in the
perpendicular configuration by a factor of $\pi R/H$, then the number
of collisions in the coplanar case is

\begin{equation}
n_\parallel\simeq 31\,n_\perp.
\end{equation}
For a RG with random orbital inclination, the number of collisions with clumps
in the disk will range between $n_\perp$ and $n_\parallel$. So as to derive
their values, we still need to estimate $t_c$.

In the standard picture of massive star formation, different parts of a
star-forming clump evolve on different timescales \citep{zin07}: the central
part collapses first due to its higher density and hence shorter free-fall
timescale.  This leads to the formation of a protostar in the core of the
clump. The outer layer contracts on a longer timescale because of its lower
density, but also due to the new source of heat at the core of the clump, the
forming protostar.

{Unlike} the standard star formation picture, in our case the clump is
optically thick. So the heat released by the protostar is kept in the clump,
and must be dissipated before the outer layer can contract further, in a
self-regulating process of the growth of the protostar and the contraction of
the outer layer. This allows us to define $t_c$. It has been shown that the
temperature of the clumps can achieve a value of the order of $T\sim10^3~K$
\citep{Bon08,map12}.  The opacity in the context of molecular clouds has been
estimated to be $\kappa\simeq0.1(T/1~{\rm K})^{1/2}~{\rm cm^2~g^{-1}}$
\citep{bell94}.  We can then calculate the optical thickness from Equation
(\ref{eqn:sigc}), $\kappa\Sigma_c\sim10^5(T/10^3~{\rm K})^{1/2}$. The
assumption of black-body in this context holds, so that the radiative cooling
rate at the surface of the outer layer is $4\pi\sigma R_c^2T^4$, where $\sigma$
is the Stefan-Boltzmann constant.

For a {\em given size} of a clump, i.e. before it can contract to a smaller
size, the outer layer will emit a {\em total} energy of $\left(4\pi\sigma
T^4R_c^2\right)t_c$. If we equate this energy with the total amount of heat
contained in the clump (i.e. in the gas and the protostar),
$GM_c^2/R_c+GM_*^2/R_*$, we have that

\begin{align}
t_c&\sim\frac{GM_*^2/R_*}{4\pi\sigma T^4R_c^2}\nonumber \\
&\sim10^5{\rm yr}\left(\frac{R}{0.1~{\rm pc}}\right)^{-2}
\left(\frac{M_*}{10^2\msun}\right)^{-2}
\left(\frac{T}{10^3~{\rm K}}\right)^{-4}.
\label{eq:tc}
\end{align}

\noindent
To relate the radius $R_*$ of the protostar to its mass $M_*$, we adopt the
empirical relation for H-burning stars that
$R_*\sim1.29\rsun(M_*/\msun)^{0.60}$ for $M_* > 1\,M_{\odot}$ and $R_*\sim
\rsun(M_*/\msun)^{0.97}$ for $M_* < 1\,M_{\odot}$ \citep[see e.g.][]{nay07}.
This is the reason why in Equation \ref{eq:tc} we have neglected the
contribution from from the gas, $GM_c^2/R_c$, since for protostars as light as
$0.2\,M_{\odot}$, the heat released is already comparable to the gravitational
energy of the gas in the clump.

Knowing that $N_c\sim10^2$ clumps with $M_c=10^2~\msun$ have formed in the
disk at $a\sim0.1$ pc, we find $n_\perp\sim2$ and $n_\parallel\sim60$,
therefore RGs with $f_{\rm loss}=2$ generally satisfy the condition $n_\perp<
n_{\rm loss}< n_\parallel$. This means a complete loss of the envelope if the
RG is in a low-inclination orbit with respect to the disk, and a partial
depletion of the envelope if the RG is in a high-inclination orbit.

There is no good reason to believe that the clumps form in a single-mass
distribution.  A more realistic one would naturally produce also lighter
clumps. This is important, because they are more efficient at removing RG
envelopes: They have higher surface densities ($\Sigma_c\propto M_c^{-1}$), and
each one contributes as many collisions with RGs as a more massive clump can
do; while the collisional cross section, $R_c^2\propto M_c^2$ is smaller, the
lifetime, $t_c\propto R_c^{-2}\propto M_c^{-2}$, is elongated. Therefore, a
disk harboring smaller clumps, of masses $M_c\sim 1-10~\msun$, could in
principle contribute significantly more to the depletion of RGs, but this
depends on their abundance, which unfortunately is not available from
observations yet.

Hence, during the self-gravitating past of the disk at the GC, a stellar core
of RGs with flat surface density distribution will be created. This core, once
formed, will last for a relaxation time.  We note that these results are in
agreement with the best fit to the observed surface density of the RGs in our
GC with an anisotropic angular-momentum distribution and a core size of $0.1$
pc \citep{mer10}.

\section{Impact on the clumps}
\label{sec.impact}

At this point one could wonder whether the accumulated impacting of RGs on to
the clumps could eventually disrupt or heat them before a successful RG
depletion.  To address this question, we estimate the amount of gas removed
from a clump after one crossing, i.e. the amount of gas ``scooped'' away in a
cylinder of height comparable to the size of the clump, $R_c$. As for the
radius of the cylinder, we note that the ratio between the radius of a RG (as
the ones considered so far) and its Bondi radius $r_{\rm B}$ is

\begin{equation}
\frac{r_*}{r_{\rm B}}\simeq \,80\, \left(\frac{R}{0.1~{\rm pc}}\right),
\label{eq.}
\end{equation}

\noindent
with $r_{\rm B} := \left(Gm_*/v_{\rm c}^2\right)$. Therefore the radius is
determined by $r_*$ and not $r_{\rm B}$. The RG {\em does} scoop away matter
from the clump because its surface density is $3-4$ orders of magnitude larger
than that of the clump. The mass loss, $\Delta m\sim r_*^2\Sigma_c$, for a
typical value of $\Sigma_c\sim10^{7-8}~\msun~{\rm pc^{-2}}$ and $r_* =
100\,R_{\odot}$, is negligible.

One could also be worried that the energy deposition could heat up the clump
and make it less dense, but this is not the case: The maximum energy that can
be deposited into a clump during each transit, $\Delta mv_c^2$, is trifling
compared to the binding energy of the clump, $GM_c^2/R_c$, since

\begin{equation}
\frac{\Delta mv_c^2}{GM_c^2/R_c}\sim\left(\frac{r_*^2}{R_cR}\right)
\left(\frac{\mbh}{M_c}\right)\sim10^{-3}
\end{equation}

\noindent
for our fiducial massive clumps.  The envelope of a RG is lost after $n_{\rm
loss}$, i.e. some 15 passages.  Such number of hits do not suffice to heat
up a clump in disk to stop star formation in it.

\section{Discussion}
\label{sec.results}

The problem of the missing bright red giants has been {the focus of an ongoing}
debate since its discovery, more than 15 years ago, by \cite{Genzel1996}. A
number of different scenarios have been invoked to explain this deficit of old
stars, but none has until now provided a simple and efficient mechanism to
solve the problem.  In this paper, {\em considering a single episode of disk
formation} at the GC, we explain the missing stars in the RGB in the natural
context of the star-forming disk that after fragmentation led to the currently
observed stellar disk in our GC. We prove with simple analytical estimates that
the distribution of clumps in the disk is sufficient to ensure the removal of
the envelopes of {the brightest} RGs.  Successive episodes of disk formation,
separated by $\sim 10^8$ yrs, based on AGN duty cycle, would have formed of the
order of ten generations of clumps at the GC.

{Toward lower luminosities, the HB stars} however have an envelope about 100
times denser {(in surface density) than those of RGB stars, as it can be easily
derived from the calculated structures of solar-metalicity of HB giants of
\cite{gir00}.  Therefore, due to momentum conservation
(Equation~\ref{eqn:sigstar}), an HB star requires on the order of 100 more
impacts with clumps to remove its envelope, although the non-linearity factor
$f_{\rm loss}$ is less clear in this case due to the lack of numerical
investigations.} We hence predict that only a low percentage of them, those
with a low inclination with respect to the disk, will have received significant
envelope damage.
Number counting of stars in the bin between 16.75 and 17.75 magnitudes in the
K-band may indicate a steepening of surface-density distribution for stars
fainter than the HB  \citep[][their figure 17]{schodel07}, pointing to the
picture of partial depletion.

We also predict that the released cores of the RGB stars populate the region of
the GC where they lost their envelopes. However, detecting these cores in
infrared (IR) surveys may be difficult: (i) the core would exhaust the
remaining hydrogen envelope in a couple of Myrs, and would hence appear as very
faint now while (ii) shifting its peak emission to shorter wavelengths,
becoming invisible in the IR filters \citep{dav05}.

To prove the densities of HB stars, we need deeper spectroscopic observations
and more complete photometric surveys down to the 18th K-magnitude.  On the
other hand, numerical simulations are required to study the effects of
non-linearity, our $n_{\rm loss}$ and $f_{\rm loss}$, in the interaction
between the clumps and the envelops of stars in the RGB but, more importantly,
of those in the HB.

\acknowledgments

This work has been supported by the Transregio 7 ``Gravitational Wave
Astronomy'' financed by the Deutsche Forschungsgemeinschaft DFG. We thank the
Kavli Institute for Theoretical Physics where one part of this work has been
completed.  This research was supported in part by the National Science
Foundation under Grant No. NSF PHY11-25915. This work has been completed at the
Al{\'a}jar
meeting\footnote{\url{http://members.aei.mpg.de/amaro-seoane/ALM13}}, and we
thank the participants for discussions, and {\'A}ngel Mill{\'a}n and Lucy
Arkwright for their hospitality at La Posada. We are indebted in particular to
Tal Alexander, Melvyn Davies, Ann-Marie Madigan, Cole Miller, and Rainer
Sch{\"o}del.

\label{lastpage}
\end{document}